\def\ROSAT{{\it ROSAT\/\ }}

\def\ASCA{{\it ASCA\/\ }}

\def\SAX{{\it SAX\/\ }}

\def\Ir{IRAS~13349+2438\ }
\def\Irc{IRAS~13349+2438}

\def\ltsima{$\; \buildrel < \over \sim \;$}
\def\simlt{\lower.5ex\hbox{\ltsima}}
\def\gtsima{$\; \buildrel > \over \sim \;$}
\def\simgt{\lower.5ex\hbox{\gtsima}}

\documentstyle[psfig]{mn}

\title[X-ray absorption by ionized oxygen in \Ir ]
{X-ray absorption by ionized oxygen in \ASCA spectra of the infrared quasar \Ir}

\author[W.N. Brandt et~al.]
{\parbox[]{6.5in}{W.N. Brandt,$^1$\thanks{Current address:
The Pennsylvania State University,
Department of Astronomy and Astrophysics,
525 Davey Lab,
University Park, Pennsylvania 16802, USA}
S. Mathur,$^1$ C.S. Reynolds$^2$
and M. Elvis$^1$}\\
\\
$^{1}$ Harvard-Smithsonian Center for Astrophysics, 60 Garden Street, 
Cambridge, Massachusetts 02138, USA\\
$^{2}$ JILA, University of Colorado, Boulder, Colorado 80309-0440, USA\\
}

\begin{document}

\maketitle

\begin{abstract}  
We present evidence for X-ray absorption by ionized 
oxygen in \ASCA spectra of the prototype infrared 
quasar \Irc. This powerful 
($L_{\rm bol}\simgt 2\times 10^{46}$~erg~s$^{-1}$)
quasar was studied in 
detail with {\it ROSAT\/}, and the combination of the
X-ray data and optical/near-infrared spectropolarimetry
strongly suggested the presence of a dusty ionized (`warm') 
absorber along the line of sight to the central X-ray source. 
The \ASCA spectra, in contrast to an earlier claim, show 
evidence for ionized oxygen edges, and the presence of
such edges appears to provide the 
most physically plausible interpretation of the data. Thus, 
the \ASCA spectra support the dusty warm absorber hypothesis.
The \ASCA data also allow the physical properties of the warm 
absorber to be constrained far better than before. A one-zone
warm absorber model indicates the ionized column 
to be in the range 
(2--6)$\times 10^{21}$ cm$^{-2}$,
and it gives an ionization parameter of 
$\xi=31^{+12}_{-12}$~erg~cm~s$^{-1}$.
The dusty warm absorber appears to 
have a density of $\simlt 3\times 10^8$ cm$^{-3}$,
and it is probably located outside the broad-line region.
The dust in the warm absorber does not appear to have been
heavily sputtered or destroyed via other means. 
Based on the \ASCA fitting, we suggest that ultraviolet absorption
lines from the warm absorber may 
be detectable and discuss how they can 
be used to further constrain the warm absorber properties.
We compare and contrast the X-ray properties of \Ir with those of
broad absorption line quasars. We comment on the steep 
$>2$~keV continuum of \Ir and examine the relevance to some
models of radiative Fe~{\sc ii} formation.
\end{abstract}


\section{Introduction} 

\Ir ($z=0.107$) is the prototype infrared quasar with
high polarization (Beichman et~al. 1986; Wills et~al. 1992,
hereafter W92). It is radio quiet and has a bolometric
luminosity of $\simgt 2\times 10^{46}$~erg~s$^{-1}$.
W92 presented a detailed model for the
optical and near-infrared
light paths in \Ir in which we are viewing the 
combination of a direct, but attenuated, quasar spectrum
and a scattered spectrum. They found an attenuation of at
least $E(B-V)=0.3$, which corresponds 
to an X-ray absorption column
density of $\simgt 1.7\times 10^{21}$ cm$^{-2}$
assuming the mean Galactic dust-to-gas ratio
(see equation 7 of Burstein \& Heiles 1978). 
Brandt, Fabian \& Pounds (1996, hereafter BFP96) used 
\ROSAT Position Sensitive Proportional Counter (PSPC)
and Wide Field Camera (WFC) data to test the W92 model. 
\Ir was very bright for the PSPC, being seen at up to
5~count~s$^{-1}$. BFP96 found large-amplitude X-ray
variability and a high X-ray luminosity which argued against the
possibility that most of the X-rays were scattered to Earth
around the attenuating matter of W92 (barring an extremely
unusual X-ray scattering `mirror'). 
Furthermore, they found that the \ROSAT spectrum constrained
the intrinsic X-ray absorption column of neutral matter to be 
about 35 times smaller than expected based on the 
optical/near-infrared 
extinction. To reconcile this large discrepancy, BFP96
invoked a dusty warm absorber, in which the
dust causing the optical extinction is embedded in ionized,
rather than neutral, gas. This reduced the expected X-ray 
absorption greatly and thereby alleviated the absorption 
discrepancy.
Fits to the \ROSAT PSPC spectra showed strong deviations
from a simple power-law model. The residuals were consistent 
with oxygen edges due to absorption by 
ionized gas, and BFP96 suggested that these 
might represent a direct detection of the dusty
warm absorber. A detailed examination of these features was 
not possible due to the limited spectral 
resolution of the PSPC.

The Japanese/USA \ASCA X-ray satellite 
(Tanaka, Inoue \& Holt 1994) offers improved 
spectral resolution over {\it ROSAT\/} that has allowed 
detailed studies of ionized oxygen edges in many Seyferts (see 
Reynolds 1997 for a recent review).
Brinkmann et~al. (1996) have recently presented
the results of an \ASCA observation of \Irc. They were forced 
to adopt the dusty warm absorber picture of BFP96 to explain 
the optical/near-infrared extinction versus cold X-ray 
column discrepancy. However, one of the main claims of
Brinkmann et~al. (1996) was that the low-energy \ASCA spectra 
ruled out the possibility of ionized oxygen edges between 
0.7--1~keV (see their sections 2.2 and 3).
Unfortunately, Brinkmann et~al. (1996) did not
provide any details of the method they used to 
show that edges were not allowed. If their claim were true, it 
would be remarkable since it 
might imply strong temporal variations of the dusty
warm absorber of \Ir (see section 3 of Brinkmann et~al. 1996).

In this paper, we carefully examine the strong claim made by
Brinkmann et~al. (1996) that the \ASCA spectra are
inconsistent with the presence of ionized oxygen edges. Our 
independent analysis suggests that ionized oxygen edges 
are, in fact, entirely consistent with the \ASCA 
data and provide a highly significant improvement in the
quality of fit. They provide a more 
natural explanation of the low energy spectral complexity
than the isolated 0.65~keV line 
suggested by Brinkmann et~al. (1996).
Warm absorbers appear to be fairly 
rare in high luminosity quasars. 
Since \Ir is one of the nearest and brightest quasars that 
shows a warm absorber, it deserves intense study 
and detailed modelling. 
We use the \ASCA data to constrain the warm absorber 
properties far better than was possible with the
\ROSAT data. We examine the dust-to-gas ratio of the 
warm absorber as well as the possibility that the
warm absorber will imprint absorption lines on 
the ultraviolet spectrum. 
We also discuss the unusually steep 2--10~keV 
continuum slope of \Ir in the context of what is seen
in other strong Fe~{\sc ii}, weak [O~{\sc iii}],
narrow-line Seyfert~1 type galaxies.

We shall adopt 
$H_0=50$ km s$^{-1}$ Mpc$^{-1}$ and 
$q_0 = \frac{1}{2}$ 
throughout. The Galactic neutral 
hydrogen column towards \Ir has been measured 
by Murphy et~al. (1996), and it is 
$(1.1\pm 0.2)\times 10^{20}$ cm$^{-2}$. 

\section{Observations and data analysis} 

\subsection{Observation details}

\Ir was observed twice with \ASCA during the AO-3 
observation round. The observations started on  
1995 June 27 and 1995 June 30. 
%
%
Both Solid-state Imaging Spectrometer 
CCD detectors (SIS0 and SIS1) and both 
Gas Imaging Spectrometer scintillation proportional 
counters (GIS2 and GIS3) were operated. 
The SIS detectors were operated in 1 CCD mode, and the 
most well-calibrated SIS chips were used (chip 1 
for SIS0 and chip 3 for SIS1). 
The lower level discriminator was not used for
the SIS detectors, and this generally facilitates
the reliable analysis of data at low energies 
(K. Mukai, private communication).
Both SIS detectors had temperatures 
in the nominal range [see section~7.2 and 
section~7.7.3 of the AO-6 \ASCA Technical 
Description (AN 97-OSS-02)].
The GIS were operated in PH mode.

We have used the `Revision~1' processed data from
Goddard Space Flight Center (GSFC) for the analysis below 
(see Day et~al. 1995b for a description of Revision~1 
processing), and data reduction was performed using {\sc ftools} 
and {\sc xselect} (see Day et~al. 1995a for a description of 
these packages and their application to \ASCA data 
analysis). 
We have used the charge transfer inefficiency (CTI) table 
released on 11 March 1997 [{\tt sisph2pi\_110397.fits};
see Dotani et~al. 1995 and section~7.7.1 of the
AO-6 \ASCA Technical Description (AN 97-OSS-02)
for discussions of CTI]. 
We have adopted the fairly strict GSFC Revision~1 
screening criteria (see Day et~al. 1995b), in order
to ensure that we are working with data of only the 
highest quality. 
%
%
After data screening, the exposure times for the first
observation are the following: 
10.9~ks for SIS0,
10.7~ks for SIS1, and 
10.8~ks each for GIS2 and GIS3.
After data screening, the exposure times for the second
observation are the following: 
6.9~ks each for SIS0 and SIS1, and
7.5~ks each for GIS2 and GIS3.  

\begin{figure*}
\centerline{\psfig{figure=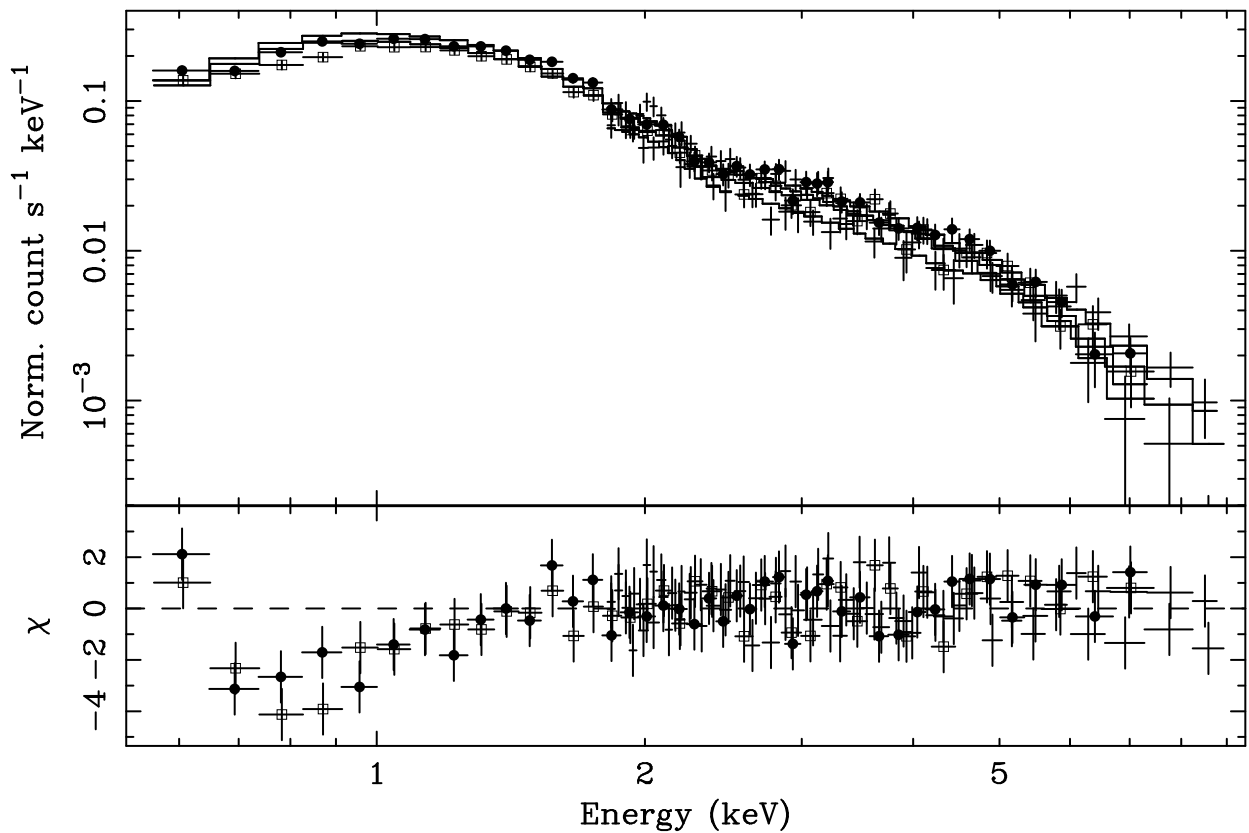,angle=270,width=1.05\textwidth}}
\caption{\ASCA SIS0 (solid dots), SIS1 (open squares) and
GIS (plain crosses) spectra of \Irc.
A power-law model has been fit to the data above 2~keV and
then extrapolated back to show the deviations from a 
power-law model at low energies.
The ordinate for the lower panel (labeled $\chi$) 
shows the fit residuals in terms of sigmas with error
bars of size one. 
Note the systematic residuals at low energies due to 
absorption edges from ionized oxygen.
This spectrum is in the observed frame, so the 0.739~keV
O~{\sc vii} K-edge threshold is redshifted to 0.668~keV. 
}
\end{figure*}

\subsection{Image and variability analyses}

Our image and variability analyses generally agree with
those described by Brinkmann et~al. (1996). We shall
therefore only describe relevant differences and
procedures directly relevant to our spectral
analysis below. 
We use circular source cells centred on 
\Ir with radii of 3.5 arcmin and 5 arcmin 
for SIS and GIS source count extraction, respectively.
Due to the position of \Ir on SIS1 chip 3, our source
cell extends slightly past the edge of the chip.
There is only a small loss of source counts, and
we have verified that this is not materially affecting
the spectral results below by using several smaller source 
cell sizes and comparing the derived results. 
We have carefully chosen local, circular, source-free
cells to use for background subtraction. We have
chosen our GIS background cells so that they lie
at similar distances from the centres of the 
detectors as \Irc. We use the counts extracted from 
these cells in all analysis below. 

We confirm the rapid X-ray variability found by 
Brinkmann et~al. (1996). The count 
rate appears to rise by about 50 per cent
in $\approx 6000$~s near the end of the first observation. 
This variability is quite rapid for a quasar with a
bolometric luminosity of $\simgt 2\times 10^{46}$ erg s$^{-1}$.
For example, if the predominant fraction of the observed
bolometric luminosity is ultimately due to accretion onto a 
Kerr black hole radiating at 20 per cent of the
Eddington limit, the variability timescale is comparable
to the light crossing timescale of the 
$8\times 10^8$ M$_\odot$ black hole. 
The required accretion efficiency, in the absence of
relativistic effects, is reasonably high ($\sim 3$ per cent) 
but not outstanding (here we use the arguments of
Fabian 1979). The rapid X-ray variability seen by
\ASCA is entirely consistent with the idea that we are seeing
X-rays directly from the quasar core (see 
section~4.2 of BFP96 for further discussion). 
 
As in Brinkmann et~al. (1996), no highly 
significant spectral variability is detected. 
Brinkmann et~al. (1996) state that the current 
\ROSAT and \ASCA data demonstrate
that \Ir varies achromatically across the entire
0.1--8~keV range. The data are indeed consistent with
this possibility. However, the \ROSAT and \ASCA 
observations were not simultaneous. It 
is still entirely possible then that
the 0.1--0.5~keV spectrum (e.g. a soft X-ray excess) varies 
independently relative to the 0.5--8~keV spectrum.

\subsection{Spectral analysis}

\subsubsection{Preparation of spectra and fitting details}

We have extracted spectra from each detector using
all of the acceptable exposure time. As no spectral
variability is evident, we have combined the
1995 June 27 and 1995 June 30 data. \Ir is
detected up to $\approx 7.5$~keV by the SIS
detectors, and it is detected up to 
$\approx 9.0$~keV by the GIS detectors.
In the rest frame of \Irc, these upper detection 
energies correspond to $\approx 8.3$~keV and
$\approx 10.0$~keV, respectively.
We have grouped the 0.55--7.5~keV SIS spectra and 
the 1--9~keV GIS spectra so that there are 12 photons
per spectral data point, to allow the use of
chi-squared fitting techniques. 
Following the suggestions of the GSFC \ASCA guest 
observer facility, we have used the {\sc sisrmg} software
to generate our SIS redistribution matrix files (rmf),
and we use the GIS rmf from 1995 March 6. We
generate our ancillary response files (arf)
using the {\sc ascaarf} software. 
We perform spectral modelling below using the X-ray 
spectral models in the {\sc xspec} spectral fitting 
package (Shafer et~al. 1991; Arnaud 1996). Unless stated
otherwise, we shall conservatively quote errors for 90.0 
per cent confidence taking all parameters to be of interest 
other than absolute normalization. 

We first performed spectral fitting for each of
the \ASCA detectors separately, and we obtained
results that were consistent to 
within the errors. We have therefore
jointly fitted the spectra from all four detectors,
and we detail these results below. In such joint 
fitting we allow the normalization for each detector
model to be free (to allow for small calibration 
uncertainties in the absolute normalizations between 
detectors; this is currently standard practice when
performing \ASCA analysis), but we tie together all other 
fit parameters across the four \ASCA detectors.   
When we quote fluxes below, we shall quote them for
the SIS0 detector. 

\subsubsection{Spectral fitting results}

We have fitted the spectra described above with a 
simple power-law model. We obtain 
$N_{\rm H}=(2.55^{+2.89}_{-2.25})\times 10^{20}$ cm$^{-2}$,
$\Gamma=2.19^{+0.08}_{-0.07}$ and
$\chi^2=543.1$ for 632 degrees of 
freedom ($\chi^2_\nu=0.859$). 
The fit is formally acceptable, although it leaves
systematic residuals in the 0.55--1.0~keV range.
The residuals are systematically above the fit 
in the $\approx$ 0.55--0.65~keV band, 
and they are systematically below the fit 
in the $\approx$ 0.65--1.0~keV band
(see below for further discussion). 
The fitted neutral absorption column is consistent with the 
Galactic value and that measured by the \ROSAT PSPC. 
Indeed, \ROSAT PSPC fitting showed the cold column
intrinsic to \Ir was $<5\times 10^{19}$ cm$^{-2}$
(see section 4.1 of BFP96). Therefore, from this
point on, we shall
adopt the Galactic neutral absorption column
unless stated otherwise. Performing a power-law fit
with the column fixed to the Galactic value, we obtain
$\Gamma=2.17^{+0.04}_{-0.04}$ and
$\chi^2=545.4$ for 633 degrees of 
freedom ($\chi^2_\nu=0.862$). 

In order to investigate the 
systematic 0.55--1.0~keV residuals further, we 
have fitted a power-law model to the data in the 
2--10~keV band (corrected for redshift) and then extrapolated this 
model down to lower energies. In the 2--10~keV band, we obtain a 
photon index of $\Gamma=2.22^{+0.08}_{-0.08}$.
The fit quality is good with $\chi^2=311.8$ for 408 degrees 
of freedom ($\chi^2_\nu=0.764$), and there are no obvious
systematic residuals in the 2--10~keV range.  
Upon extrapolation of this model to lower energies, negative
residuals are clearly apparent 
in the 0.65--1.3~keV band (see Figure 1).
Such residuals are frequently seen in \ASCA data when 
ionized absorbing matter lies along the line of sight
(compare our Figure~1 with figure~3 of Reynolds 1997),
and they are attributed to 
the presence of edges due to ionized oxygen 
in the X-ray spectrum. In order to examine possible 
oxygen absorption in the spectrum of \Irc, we 
have first added an edge to the power-law model from 
the previous paragraph. We obtain 
$\Gamma=2.27^{+0.07}_{-0.07}$,
$E_{\rm Edge}=0.75^{+0.04}_{-0.04}$ keV,
$\tau_{\rm Edge}=0.41^{+0.14}_{-0.13}$ and
$\chi^2=517.4$ for 631 degrees of freedom 
($\chi^2_\nu=0.819$; the edge 
parameter errors are quoted for
$\Delta \chi^2=2.71$ and the edge threshold
energy has been corrected for redshift).
The fit quality is good, and the low energy systematic 
residuals described in the previous paragraph are removed. 
We show confidence contours for the edge parameters
in Figure~2. 
The edge energies usually seen in 
Seyfert ionized absorbers are 
0.739~keV for the O~{\sc vii} K edge and
0.871~keV for the O~{\sc viii} K edge.
Our best fitting edge energy lies between these two values.
It is in good agreement with the O~{\sc vii} K edge energy,
which is the strongest edge usually seen in warm absorbers.
Applying the $F$-test
(see tables C-5 and C-6 of Bevington \& Robinson 1992) with 
$\Delta \chi^2=-28.0$ for the 
two additional edge parameters, we find 
that the addition of the edge makes a significant improvement 
to the fit quality with over 99 per cent confidence. 
We have tried adding edges at other arbitrary energies
from 1.0--2.5~keV, and none reduces the value of $\chi^2$
enough that it makes a highly significant improvement to the
fit quality (this helps to validate the significance of
our $F$-test results).

We have also modelled the residuals using
two edges with their threshold energies fixed at 
the 0.739~keV (Edge~1) and 0.871~keV (Edge~2) 
threshold energies for O~{\sc vii} and O~{\sc viii}.
We obtain 
$\Gamma=2.28^{+0.07}_{-0.07}$,
$\tau_{\rm Edge\ 1}=0.37^{+0.15}_{-0.15}$,
$\tau_{\rm Edge\ 2}=0.09^{+0.11}_{-0.09}$ and
$\chi^2=515.8$ for 631 degrees of freedom 
($\chi^2_\nu=0.817$; the errors for the edge
depths are for $\Delta\chi^2=2.71$).
Applying the $F$-test with $\Delta \chi^2=-29.6$ for 
the two additional edge parameters, we again
find a highly significant improvement with
over 99 per cent confidence (as compared to the
simple power-law model). We cannot statistically 
discriminate between the one-edge model of the
previous paragraph and the two-edge model 
discussed here. Based on \ASCA observations of
nearby Seyferts, we suspect that the two-edge
model is the most physically appropriate.

We have verified that potential calibration uncertainties 
at low energies do not appear to be affecting the basic 
nature of our results. Dotani et~al. (1996) have
examined calibration systematics in this energy range and
find that they are $\approx 10$ per cent or less (see
their section~2.3). We comment
that we are using the data in the 0.55--0.6~keV band only 
for the purpose of defining the basic continuum level below 
the edge. This use of the 0.55--0.6~keV data requires 
somewhat less of the calibration than does, for 
example, precision X-ray emission line work in this 
energy range (see George, Turner \& Netzer 1995 who examine
a possible O~{\sc vii} emission line at 0.57~keV). 
If we completely ignore the data in the 0.55--0.6~keV band 
we obtain the same qualitative results. In this case, the 
addition of two edges with their threshold energies fixed
at 0.739~keV and 0.871~keV gives $\Delta\chi^2=-21.1$ for 
two additional parameters (relative to a simple power-law
model). The $F$-test indicates 
that this is a significant improvement in $\chi^2$ with 
over 99 per cent confidence. 

\begin{figure}
\centerline{\psfig{figure=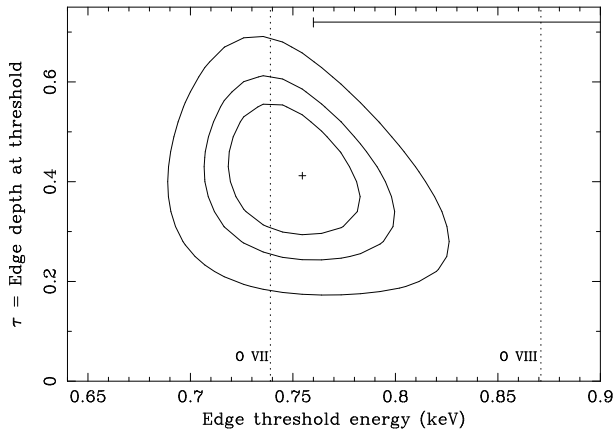,angle=270,width=0.5\textwidth}}
\caption{\ASCA confidence contours for the edge parameters.
Contour levels are for 
$\Delta\chi^2=2.30$ (inner contour; 68.3 per cent confidence
for two parameters of interest), 
$\Delta\chi^2=4.61$ (middle contour; 90.0 per cent confidence
for two parameters of interest) and
$\Delta\chi^2=9.21$ (outer contour; 99.0 per cent confidence
for two parameters of interest).
The fitted edge energy has been 
corrected for cosmological redshift.
The threshold energies for the O~{\sc vii} K edge (0.739~keV)
and the O~{\sc viii} K edge (0.871~keV) are shown as vertical 
dotted lines. 
The lower part of the edge threshold 
energy range derived from \ROSAT data is
shown as the horizontal solid line at the top of the figure
(the y-axis position of the line is arbitrary). 
Note that the derived edge energy from 
the \ASCA fitting is in good agreement with
with the threshold energy for the O~{\sc vii} K edge. This is
the strongest edge usually seen in Seyfert warm absorbers.
}
\end{figure}

To model our data more realistically, we have fitted the \ASCA 
spectrum with a one-zone warm gas absorption model constructed
from {\sc cloudy} (Ferland 1992) calculations. The model is
the same as that described in section 3.1 of 
Reynolds et~al. (1995), and line emission from the
warm absorber is consistently included assuming a 
covering fraction of unity for the warm absorber
(even for unity covering fraction, the lines are
quite small for the range of physical parameters under 
consideration here and do not substantially affect
the fitting given the statistics of these \ASCA data).
This model gives a good fit with $\chi^2=518.6$ for 631 
degrees of freedom. The derived fit parameters are
$\Gamma=2.31^{+0.07}_{-0.15}$,
log($N_{\rm H,warm})=21.59^{+0.15}_{-0.26}$ and
$\xi=31.4^{+12.1}_{-12.2}$~erg~cm~s$^{-1}$
(warm absorber parameter errors are for $\Delta \chi^2=2.71$).
The ionization parameter, $\xi$, is defined to be 
$L/(n R^2)$ where $L$ is the ionizing luminosity,
$n$ is the number density of the absorbing plasma and
$R$ is the distance to the illuminating source. 
Relative to the simple power-law model we obtain 
$\Delta\chi^2=-26.8$. The $F$-test indicates that 
the one-zone warm absorber model makes a significant
improvement in fit quality 
with over 99 per cent confidence.
The fit quality is not quite as good as
that for just two simple edges, although there is
no statistically significant difference.

We have used the power-law and one-zone warm absorber 
model to compute mean fluxes and the mean luminosity 
in the 2--10~keV band.
The absorbed and absorption-corrected 
2--10~keV fluxes are both
$5.7\times 10^{-12}$~erg~cm$^{-2}$~s$^{-1}$,
and the derived 2--10~keV isotropic luminosity is 
$3.0\times 10^{44}$~erg~s$^{-1}$.
We have also used the same model to compute mean
fluxes and the mean luminosity in the 0.55--2~keV band.
The absorbed and absorption-corrected 
0.55--2~keV fluxes are 
$6.0\times 10^{-12}$ erg cm$^{-2}$ s$^{-1}$
and 
$7.0\times 10^{-12}$ erg cm$^{-2}$ s$^{-1}$, respectively.
The derived 0.55--2~keV isotropic luminosity is 
$3.6\times 10^{44}$ erg s$^{-1}$.
Fluxes and luminosities derived using the models with 
simple edges described above are very similar 
to those quoted here. 

Finally, we have fitted the 3--10~keV data (corrected for
redshift) with a simple power-law model to look for any 
evidence of continuum curvature. We obtain an acceptable 
fit and a photon
index of $2.20^{+0.16}_{-0.17}$. There is no
significant evidence for continuum curvature
(i.e. the 2--10~keV and 3--10~keV photon indices are
consistent with each other) although the 
constraints are not tight. 

\subsubsection{Safety checks}

Due to the fact that our results regarding the 
presence of oxygen edges are in contradiction
to the claims of Brinkmann et~al. (1996), we have repeated
our analysis in several ways to ensure its reliability. Data
analysis has been performed independently by three of the
authors (WNB, SM, CSR) using \ASCA software installations
at two different institutions, and all results are in generally
good agreement. 
We have repeated our analysis using the 1994 November 9 SIS rmf 
files, and using these does not materially change our conclusions.
We have also repeated our analysis using the GSFC blank sky 
fields for background subtraction, and the use of these 
does not change our conclusions. 
If we use older CTI tables (see Section 2.1) the agreement
between SIS0 and SIS1 is not as good, but oxygen edges
still provide a reduction in $\chi^2$ that is significant
with over 99 per cent confidence.  

Additional spectral fitting by Y. Ogasaka (who was primarily 
responsible for the \ASCA analysis presented in Brinkmann et~al. 1996)
now also shows that oxygen edges are entirely consistent with the 
actual Brinkmann et~al. (1996) spectra (Y. Ogasaka, private
communication).

\section{Discussion} 

\subsection{Spectral features from warm absorption} 

\subsubsection{Oxygen edges in the ASCA spectra}

We find that the presence of 
ionized oxygen edges is entirely
consistent with the \ASCA spectra of \Irc. 
The addition of ionized oxygen 
absorption provides a highly significant
improvement in $\chi^2$, and the derived energy range for 
the absorption comes out to be in good agreement with that
seen from O~{\sc vii} and O~{\sc viii} in many other Seyfert~1
class objects. While we cannot statistically prove that
oxygen edges are the only possible interpretation of the
low-energy residuals (a fairly common situation in X-ray data
analysis), such absorption is expected due to the
large optical/near-infrared extinction versus cold X-ray column
discrepancy established for \Ir (BFP96). An
{\it isolated\/} oxygen emission line at 0.65~keV, as proposed
by Brinkmann et~al. (1996), has not been seen in other
Seyfert~1 class objects to our knowledge (\Ir should be treated 
as a Seyfert~1 class object in this context due to the fact that 
its X-ray variability demonstrates that we are seeing the
X-ray continuum directly from its black hole region). 
While low-energy oxygen emission lines have been claimed in 
some Seyfert~1s (e.g. George et~al. 1995), they are 
thought to be associated with warm absorbers and have
been seen to be accompanied by much stronger oxygen edge 
features. An isolated oxygen line could perhaps arise
if there were line-emitting warm absorber type gas out of 
the line of sight but not along the line of sight. However,
as noted above, we have good, independent reasons 
to think that a warm absorber in \Ir lies
along the line of sight. While, of course, some 
contribution to the \ASCA residuals is probably due to 
oxygen line emission (indeed, this is required by atomic 
physics provided the ionized gas subtends a non-negligible
solid angle), it seems most plausible that the dominant contribution
is due to oxygen absorption edges. It is difficult to pinpoint the
precise origin of our disagreement with 
Brinkmann et~al. (1996), as they present 
insufficient details of the
analysis behind their strong claim that edges in the 
0.7--1.0~keV range are inconsistent with the \ASCA data. 
Instead we have established that
oxygen edges are in good statistical agreement with
the \ASCA data and moreover provide a physically 
plausible explanation for the observed residuals.

\subsubsection{ASCA warm absorber constraints}

The \ASCA data provide much more reliable warm absorber constraints 
than the \ROSAT data. The limited spectral resolution of \ROSAT can 
make fitting constraints on warm absorbers difficult to obtain. 
\ROSAT fitting with an edge model
gave a rather high edge energy of $0.86^{+0.10}_{-0.10}$~keV
although the statistical error bar was also large (compare
with Figure~2). BFP96 clearly discussed 
the difficulties of precision modelling of the
warm absorber using \ROSAT data. In particular, there is 
probably a soft excess in the low-energy end of 
the \ROSAT band which affects the fitted \ROSAT warm 
absorber parameters due to the limited spectral resolution 
of the PSPC (see section 3.3 of BFP96). 

Fitting the one-zone warm absorption model to the 
\ASCA data indicates the ionized column along the line
of sight is $\approx$(2--6)$\times 10^{21}$ cm$^{-2}$, which 
is rather similar to that seen in many less luminous
Seyferts. The fitted \ASCA warm column is smaller than
the \ROSAT warm column of 
log($N_{\rm H,warm})=22.78^{+1.20}_{-1.18}$, although the 
two measured values are not formally discrepant. The 
\ASCA errors on the column are about 4 times smaller than the
\ROSAT errors. 
W92 argue the extinction to the nucleus of \Ir has
$E(B-V)$ in the range 0.3--0.7 (see their section 4.4.3). 
Using the mean Galactic dust-to-gas ratio, the gas 
column range that corresponds to this $E(B-V)$ range is
(1.7--4.0)$\times 10^{21}$ cm$^{-2}$.
The good agreement of the 
fitted \ASCA warm column range with 
this range is remarkable, given the very different environments
under consideration (i.e. the local interstellar medium as
compared to ionized gas around a quasar).
Provided the one-zone warm absorption
modelling is roughly appropriate, it appears likely that the
dust in the warm absorber of \Ir has not been heavily
sputtered or destroyed via other means. 
Given the apparent lack of heavy 
sputtering, equation 44 of Draine \& Salpeter (1979)
suggests that the temperature of the ionized gas must be less than
about $10^6$~K provided its density is $\simgt 5\times 10^4$~cm$^{-3}$
(also see section VI of Burke \& Silk 1974, who quote a somewhat
lower sputtering threshold temperature). Collisionally
ionized gas with O~{\sc vii} and O~{\sc viii} would have a
temperature of $\simgt 10^6$~K (see table 3 of 
Shull \& Van Steenberg 1982), so it seems unlikely that
the warm absorber gas is collisionally ionized. This
appears to vindicate our implicit 
assumption of photoionization.

The \ASCA ionization parameter is 4.6 times lower than 
and inconsistent with the value of
$\xi=145^{+80}_{-56}$ erg~cm~s$^{-1}$ derived from the
\ROSAT data. The \ASCA error bars are again much
smaller than the \ROSAT error bars. The \ASCA ionization 
parameter is in better agreement with what is usually
seen in Seyfert~1 class objects (compare with table 6 of 
Reynolds 1997). Adding a soft excess to the \ROSAT data 
(see above) allows 
for consistency between the \ROSAT and \ASCA warm absorber
parameters, although the \ROSAT parameters are very
poorly constrained. In addition, of course, there may well
have been significant warm absorber variability in the
$\sim 3$ years between the \ROSAT and \ASCA observations.

Given the ionization parameter of the warm absorber and
the fact that it contains dust, we can derive an upper 
limit on its density. We have the relation
$n=L/\xi R^2$ (see Section 2.3.2 for the definitions of
the symbols). From the \ASCA fitting we have
$\xi\approx 31$~erg~cm~s$^{-1}$, and from the
observed \ROSAT and \ASCA luminosities we derive a mean
ionizing luminosity of $\approx 1.5\times 10^{45}$ erg~s$^{-1}$. 
In order to avoid sublimation the dust must be
$\simgt 4\times 10^{17}$ cm from the nucleus (see 
section~5.3 of Laor \& Draine 1993). Thus we derive 
$n\simlt 3\times 10^8$ cm$^{-3}$, which is significantly
less than the density of a typical cloud in the 
broad-line region. 
The dusty warm absorber of \Ir is probably located outside 
its broad-line region, perhaps in its intermediate-line 
region or narrow-line region.
We also comment that if the warm
absorber gas is above the dust sputtering threshold 
temperature, then the apparent 
lack of heavy sputtering can be used
to derive stringent constraints on the warm absorber
density (see equation 35 of Burke \& Silk 1974 and
equation 44 of Draine \& Salpeter 1979).

An observation with {\it AXAF\/}, or even a longer
observation with \ASCA (note the total exposure time 
for these two observations was rather short), would allow
further constraints to be placed on the dusty warm absorber. 
It would be especially interesting to study warm absorber
variability and to better constrain the radial velocity
of the X-ray absorbing gas. The agreement between the
fitted \ASCA edge energy and the O~{\sc vii} K-edge 
threshold energy suggests that the warm absorber
is moving at less than about 6 per cent of the speed
of light.

\subsubsection{UV absorption lines --- constraints, predictions
and relevance to broad absorption line quasars}

In several active galaxies, ionized X-ray absorption may well
be associated with ultraviolet absorption lines 
(e.g. Mathur, Elvis \& Wilkes 1995 and references therein). 
To investigate whether this is the case for \Irc, we have
looked for ultraviolet absorption lines in the 
{\it International Ultraviolet Explorer (IUE)\/} spectrum 
from Lanzetta, Turnshek \& Sandoval (1993). No
strong lines are seen in the spectrum from the SWP
instrument, but it is fairly noisy and strong line
constraints are not available from these data. The
absorption feature at $\approx 3050$~\AA\ 
near the Mg~{\sc ii} emission line in the
spectrum from the LWP instrument is  
most probably an instrumental reseaux 
mark (see figure 2 of Kolman et~al. 1992). 

Ionization fractions of ions ($f_{\rm ion}$) like 
O~{\sc vi}, 
C~{\sc iv} and 
N~{\sc v} can be
determined for a given ionization parameter (see figure 4 in
Mathur, Wilkes and Aldcroft 1997 for $f_{\rm ion}$ as a function
of ionization parameter). The best-fit value of the ionization parameter
of the warm absorber in \Ir is $\xi=31.4$ (corresponding to
$U\sim 1.3$
for a standard active galactic nucleus continuum as 
defined in {\sc cloudy}). The absorbing column
density in each ion is then 
$N_{\rm ion}= N_{\rm H, warm} \times f_{\rm ion} \times A_{\rm element}$ 
where $A_{\rm element}$ is the abundance of the element relative to hydrogen.
In the warm absorber of \Irc, the C~{\sc iv} and N~{\sc v}  
absorption line column densities are  expected to be
$\simgt 10^{16}$ cm$^{-2}$ while the column density in O~{\sc vi} would be
$\simgt 7\times10^{17}$ cm$^{-2}$. We thus expect to see high ionization
absorption lines of 
O~{\sc vi}, 
C~{\sc iv},
N~{\sc v} and
Ly$\alpha$
in high quality ultraviolet spectra of \Irc.
Note, however, that if the
ionization parameter is towards the higher end of the
range allowed by the \ASCA fitting, the 
expected column densities would be
significantly lower. Spectra from the 
{\it Hubble Space Telescope\/} would allow the ionization 
state of the dusty warm absorber 
to be further constrained. If lines are detected,
they might also constrain its velocity along the
line of sight. 

High-quality ultraviolet spectra are also important to
obtain for \Ir due to its high polarization and its
radio-quiet nature. Almost all highly-polarized, 
radio-quiet quasars are thought to be 
broad absorption line (BAL) quasars, and so it would be 
interesting to know if \Ir is a BAL quasar as well.
\Ir has the strong optical Fe~{\sc ii} emission, weak
[O~{\sc iii}] emission and strong far-infrared
emission that are typical of low-ionization
BAL quasars (e.g. section~3.2 of
Boroson \& Meyers 1992).
BAL quasars are generally known 
to be weak soft X-ray emitters,
and this is probably due to heavy X-ray 
absorption in their nuclei
(see Green \& Mathur 1996 and references therein).
Their small soft X-ray fluxes have made them
elusive X-ray sources, and to date there is no
confirmed BAL quasar with a high-quality X-ray
spectrum (i.e. one that allows both the column density
and ionization state of the absorbing matter to be
determined with reasonable precision by X-ray
spectral fitting).
In sharp contrast to the general situation
for BAL quasars, \Ir is
a strong soft X-ray source and was one of the
brightest active galaxies in the \ROSAT sky.
It appears to have much less X-ray absorption than 
is often seen in BAL quasars. If ultraviolet
spectra do show \Ir to be a BAL quasar, then
it will be the first BAL quasar with a high-quality
X-ray spectrum.
\Ir might represent an intermediate object between 
BAL quasars and other quasars with associated 
absorption systems. It may be oriented such that 
our line of sight to the central X-ray source only 
skims the surface of the BAL outflow. We might then
only be looking through an ionized surface layer
of the outflow. 
Finally, some models for BAL outflows suggest that
they are accelerated by radiation pressure on 
dust (see section 3.3 of de Kool 1997 for a review).
The fact that we find evidence for dusty, ionized
gas in \Ir may lend observational support to such
models.

\subsection{The steep hard X-ray continuum } 

The \ASCA data show the 2--10~keV spectrum of \Ir to be
unusually steep with $\Gamma\approx 2.22$. Comparing this
photon index with the photon index distributions shown
in figure~10 of Nandra et~al. (1997) and figure~2 of
Reynolds (1997) makes this clear, as \Ir is steeper than
all of these objects except for Mrk~766. Mrk~766 is a
narrow-line Seyfert~1 galaxy (see Osterbrock \& Pogge 1985; 
Goodrich 1989; Boller, Brandt \& Fink 1996) 
that has been claimed to have strong
photon index variability by Leighly et~al. (1996).

BFP96 compared the properties of \Ir to those seen in 
many narrow-line Seyfert~1s, and they found a striking
number of similarities including 
strong Fe~{\sc ii}, 
weak [O~{\sc iii}], 
relatively narrow permitted line cores
(with FWHM of $\approx 2100$ km s$^{-1}$), 
a relatively soft X-ray spectrum
and rapid X-ray variability for an
object of its luminosity (see their section 4.5
and compare with the primary eigenvector of 
Boroson \& Green 1992). 
Brandt, Mathur \& Elvis (1997) have systematically compared 
the 2--10~keV \ASCA power-law slopes of soft \ROSAT 
narrow-line Seyfert~1s to those of Seyfert~1s with larger
H$\beta$ FWHM
values. They found that 
soft \ROSAT narrow-line Seyfert~1s as a class appear to
have generally steeper 2--10~keV power-law slopes (they also 
pointed out that some soft \ROSAT narrow-line Seyfert~1s  
may have interesting 2--10~keV continuum curvature). \Ir 
appears to follow the hard photon index trend seen in many
narrow-line Seyfert~1s, further strengthening the 
comparisons made in section 4.5 of BFP96. In addition,
the rapid X-ray variability seen by \ASCA adds to the
similarity.

The steep 2--10~keV slopes of \Ir and some other
strong optical Fe~{\sc ii} emitters (e.g. I~Zwicky~1) in
Brandt et~al. (1997) have implications for
models of radiative Fe~{\sc ii} formation.
While some narrow-line Seyfert~1s may have interesting
2--10~keV continuum 
curvature (see Brandt et~al., in preparation),
none has shown evidence for a spectral flattening
to a slope less than those of typical Seyfert 1s.
Therefore, in the absence of any strong spectral
flattening above the \ASCA band, their hard X-ray 
emission will be comparatively weak. 
This provides evidence 
against models of optical Fe~{\sc ii} line formation that
require very flat hard X-ray spectra and strong hard X-ray
emission (e.g. the Compton-heating model of 
Collin-Souffrin, Hameury \& Joly 1988; see the
last paragraph of their section 4). While this
does not settle the debate as to whether optical 
Fe~{\sc ii} is formed in a mechanically heated or
radiatively heated medium, it does argue against one
radiative heating model. 
Other radiative heating models predict an 
anticorrelation between Fe~{\sc ii}/H$\beta$ and
hard X-ray luminosity (see section~4 of Joly 1993
and references therein), and this appears to be
what the currently available X-ray data suggest. 
Further \ASCA and \SAX
measurements of the hard X-ray 
spectra of strong optical Fe~{\sc ii}
emitters (e.g. PHL~1092, Mrk~42 and Mrk~957) will provide
additional constraints, and such observations are
planned.

\section{Summary} 

We have analysed and interpreted the \ASCA data for the
prototype infrared quasar \Irc. Our main results are
the following:

(1) The \ASCA spectra show systematic deviations from a 
power-law model in the 0.55--1~keV band. The addition
of K edges from O~{\sc vii} and O~{\sc viii} removes
the systematic deviations and provides a highly 
significant improvement in fit quality. The \ASCA
evidence for oxygen absorption is consistent with the 
dusty warm absorber hypothesis put forward based 
on \ROSAT and optical/near-infrared data. 

(2) The \ASCA data greatly improve the X-ray constraints
on the warm absorber. Fitting a one-zone warm absorber
model gives an ionized column of
(2--6)$\times 10^{21}$ cm$^{-2}$
and an ionization parameter of 
$\xi=31^{+12}_{-12}$~erg~cm~s$^{-1}$.
Assuming the mean Galactic dust-to-gas ratio,
the ionized column and the column expected from
the $E(B-V)$ measurement are in good agreement.
The dusty warm absorber is probably located 
outside the broad-line region and has a 
number density $\simlt 3\times 10^8$~cm$^{-3}$.

(3) We use the \ASCA data to make predictions about
ultraviolet absorption in the context of the unified
ultraviolet/X-ray absorber model. Many properties of
\Ir are similar to those seen in 
broad absorption line quasars, but in
contrast to many broad absorption line 
quasars \Ir is bright in the soft
X-ray band. If \Ir is a broad absorption line 
quasar, then it is the first
one with a high-quality X-ray spectrum.

(4) The 2--10~keV spectrum of \Ir is steep
($\Gamma=2.22^{+0.08}_{-0.08}$). This and other
properties of \Ir resemble those seen in narrow-line
Seyfert~1 galaxies. We discuss the relevance of the steep
2--10~keV slope to some models of radiative Fe~{\sc ii}
formation.

\section*{Acknowledgments}

We gratefully acknowledge financial support from the 
Smithsonian Institution (WNB),
NASA grant NAG5-3249 (SM),
NSF grant AST-9529175 (CSR) and
NASA grant NAG5-3066 (ME). 
We thank 
J.P. Halpern,
Y. Ogasaka, 
B.J. Wilkes and
B.J. Wills for 
helpful discussions. We thank K. Mukai
and R.F. Mushotzky for advice regarding
\ASCA calibration issues. 
We thank the members of the \ASCA team who have made 
these observations possible.

\bsp


\begin{thebibliography}{}

\bibitem []{} Arnaud K.A., 1996, in Jacoby G., Barnes J., eds, 
Astronomical Data Analysis Software and Systems V: 
ASP Conference Series \# 101. ASP Press, San 
Francisco, p. 17

\bibitem []{} Beichman C.A., Soifer B.T., Helou G., Chester T.J., 
Neugebauer G., Gillett F.C., Low F.J., 1986, ApJ, 308, L1

\bibitem []{} Bevington P.R., Robinson D.K., 1992, 
Data Reduction and Error Analysis for the Physical Sciences:
Second Edition. McGraw Hill, New York

\bibitem []{} Boller Th., Brandt W.N., Fink H., 
1996, A\&A, 305, 53 (BBF96)

\bibitem []{} Boroson T.A., Green R.F., 1992, ApJS, 80, 109

\bibitem []{} Boroson T.A., Meyers K.A., 1992, ApJ, 397, 442

\bibitem []{} Brandt W.N, Fabian A.C., Pounds K.A., 1996, 
MNRAS, 278, 326 (BFP96) 

\bibitem []{} Brandt W.N., Mathur S., Elvis M., 1997, 
MNRAS, 285, L25

\bibitem []{} Brinkmann W., Kawai N., Ogasaka Y., Siebert J., 
1996, A\&A, 316, L9

\bibitem []{} Burke J.R., Silk J., 1974, ApJ, 190, 1

\bibitem []{} Burstein D., Heiles C., 1978, ApJ, 225, 40 

\bibitem []{} Collin-Souffrin S., Hameury J.M., Joly M.,
1988, A\&A, 205, 19

\bibitem []{} Day C., Arnaud K., Ebisawa K., 
Gotthelf E., Ingham J., Mukai K., White N., 1995a, 
The ABC Guide to \ASCA Data Reduction: version 4, NASA/GSFC

\bibitem []{} Day C., Jennings D., Seufert E., Watkins R., 1995b, 
\ASCA Getting Started Guide for Revision 1 Data: version 4.1, NASA/GSFC

\bibitem []{} de Kool M., 1997, in Weymann R., Shlosman I., Arav N.,
eds, Mass Ejection from AGN. WSA Press, Pasadena, in press

\bibitem []{} Dotani T., Yamashita A., Rasmussen A., Team S., 
1995, ASCA News, 3, 25

\bibitem []{} Dotani T., et~al., 1996, ASCA News, 4, 3

\bibitem []{} Draine B.T., Salpeter E.E., 1979, ApJ, 231, 77

\bibitem []{} Fabian A.C., 1979, Proc. Roy. Soc. London A, 366, 449

\bibitem []{} Ferland G.J., 1992, University of Kentucky Department of
Physics and Astronomy Internal Report

\bibitem []{} George I.M., Turner T.J., Netzer H., 1995, 
ApJ, 438, L67

\bibitem []{} Goodrich R.W., 1989, ApJ, 342, 224

\bibitem []{} Gotthelf E., 1996, \ASCA News, 4, 31

\bibitem []{} Green P.J., Mathur S., 1996, ApJ, 462, 637

\bibitem []{} Joly M., 1993, Ann. Phys. Fr., 18, 241

\bibitem []{} Kolman M., Halpern J.P., Shrader C.R., 
Filippenko A.V., Fink H.H., Schaeidt S.G., 1993, ApJ, 402, 514 

\bibitem []{} Lanzetta K.M., Turnshek D.A., Sandoval J., 1993, 
ApJS, 84, 109

\bibitem []{} Laor A., Draine B.T., 1993, ApJ, 402, 441

\bibitem []{} Laor A., Fiore F., Elvis M., Wilkes B.J.,
McDowell J.C., 1997, ApJ, 477, 93

\bibitem []{} Leighly K.M., Mushotzky R.F., Yaqoob T., Kunieda H.,
Edelson R., 1996, ApJ, 469, 14

\bibitem []{} Mathur S., Elvis M., Wilkes B.J., 1995, ApJ, 452, 230

\bibitem []{} Mathur S.,  Wilkes B.J., Aldcroft T., 1997, ApJ, 478, 182 

\bibitem []{} Murphy E.M., Lockman F.J., Laor A., Elvis M., 1996,
ApJS, 105, 369

\bibitem []{} Nandra K., George I.M., Mushotzky R.F., Turner T.J., 
Yaqoob T., 1997, ApJ, 477, 602 

\bibitem []{} Osterbrock D.E., Pogge R.W., 1985, ApJ, 297, 166

\bibitem []{} Pounds K.A., Brandt W.N., 1997, 
in Makino F., Mitsuda K., eds,
X-Ray Imaging and Spectroscopy of Cosmic Hot Plasmas: \ASCA Third 
Anniversary Proceedings. Univ. Acad. Press, Tokyo, p. 209

\bibitem []{} Reynolds C.S., Fabian A.C., Nandra K., Inoue H., 
Kunieda H., Iwasawa K., 1995, MNRAS, 277, 901

\bibitem []{} Reynolds C.S., 1997, MNRAS, 286, 513

\bibitem []{} Shafer R.A., Haberl F., Arnaud K.A., Tennant A.F., 1991,
XSPEC Users Guide. ESA Publications, Noordwijk

\bibitem []{} Shull J.M., Van Steenberg M., 1982, ApJS, 48, 95

\bibitem []{} Tanaka Y., Inoue H., Holt S.S., 1994, PASJ, 46, L37

\bibitem []{} Wills B.J., Wills D., Evans N.J., Natta A., Thompson
K.L., Breger M., Sitko M.L., 1992, ApJ, 400, 96 (W92) 


\end{thebibliography}
\end{document}